\documentclass[a4paper]{jpconf}
\usepackage{graphicx}
\begin{document}
\title{Persistent and Reversible Phase Control in GdMnO$_3$ near the Phase Boundary}

\author{H. Kuwahara, M. Akaki, J. Tozawa, M. Hitomi, K. Noda,\linebreak and D. Akahoshi}

\address{Department of Physics, Sophia University, Tokyo 102-8554, Japan}

\ead{h-kuwaha@sophia.ac.jp}

\begin{abstract}
We have investigated temperature and magnetic-field dependence of dielectric properties in the orthorhombic GdMnO$_3$ single crystal which is located near the phase boundary between the ferroelectric/spiral-antiferromagnetic phase and the paraelectric/$A$-type-antiferromagnetic one. In this compound, strong phase competition between these two phases results in a unique phase diagram with large temperature and magnetic-field hystereses.  Based on the phase diagram, we have successfully demonstrated the persistent and reversible phase switching between them by application of magnetic fields.
    
\end{abstract}

\section{Introduction}
Since the discovery of unusual gigantic magnetoelectric response in orthorhombic TbMnO$_3$, a renewed interest has been attracted in magnetoelectric or multiferroic materials from the viewpoint of future device applications such as magnetic- (electric-) field controlled ferroelectric (ferromagnetic) memories or transducers.
In this work, we have focused multiferroic GdMnO$_3$, because it is located between TbMnO$_3$ showing the the \underline{f}erroelectric/\underline{s}piral-antiferromagnetic (FS) phase and EuMnO$_3$ showing the \underline{p}araelectric/\underline{$A$}-type-antiferromagnetic (PA) one~\cite{Arima,Goto,Kuwahara}.  We have simply expected that orthorhombic GdMnO$_3$ near the phase boundary is suitable to accomplish the phase switching between the FS and PA phases.  For the purpose of realizing the phase switching, we have systematically investigated temperature ($T$) and magnetic-field ($H$) dependence of dielectric properties of GdMnO$_3$ single crystal in external magnetic fields along the crystal axes.
As a result, we have obtained the detailed magnetoelectric phase diagram of GdMnO$_3$ in the $H$-$T$ plane, which is quite complex because there exists a metastable state (hysteretic region) due to the strong phase competition between the FS and PA ones.
In this paper, we present a novel persistent and reversible phase control between them by application of magnetic fields.
  

\section{Experiment}

The single crystalline sample was grown by the floating zone method.  We performed x-ray-diffraction measurement on the resulting crystal at room temperature, and confirmed that the samples have an orthorhombic {\it Pbnm} structure without any impurity phases.  All specimens used in this study were cut along the crystallographic principal axes into a rectangular shape by means of back-reflection Laue technique.  Measurements of temperature and magnetic-field dependence of the dielectric constant and the spontaneous ferroelectric polarization were carried out in a temperature-controllable cryostat with a superconducting magnet that provided a field up to 8~T\@.  The dielectric constant was measured at 100~kHz with an $LCR$ meter (Agilent, 4284~A).  The spontaneous polarization was obtained by the accumulation of a pyroelectric current for temperature scans and a displacement current for magnetic-field ones.  

\section{Results and Discussion}

\begin{figure}[h]

\includegraphics[width=160mm, keepaspectratio, clip]{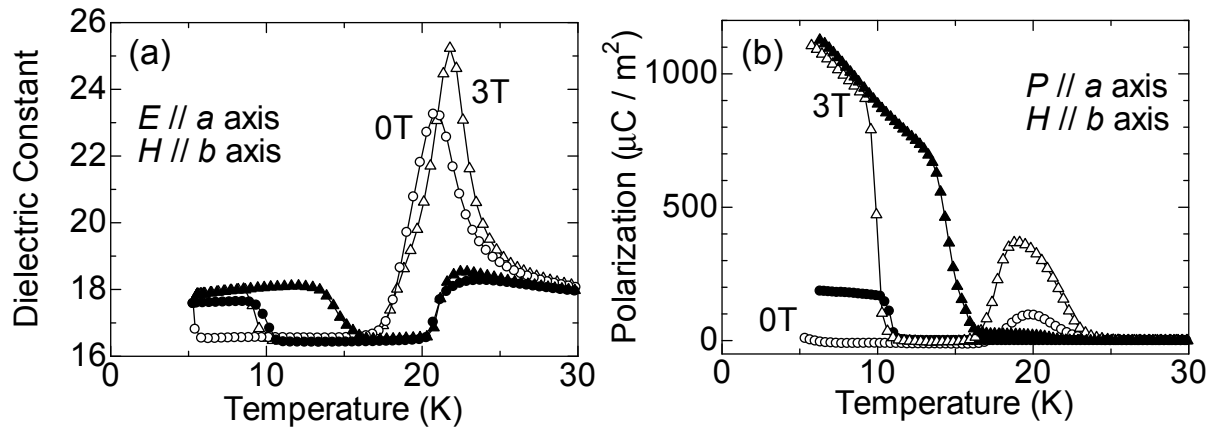}
\caption{\label{fig1}Temperature dependence of dielectric constant $\varepsilon_a$ (a) and ferroelectric polarization $P_a$ (b) along the $a$ axis for GdMnO$_3$ in a zero magnetic field (circle) and a field of 3 T (triangle).  Magnetic field was applied along the $b$ axis.  Open and solid symbols mean the cooling and warming scans, respectively.}

\end{figure}

\noindent First, we show in Fig.~1 temperature dependence of the dielectric constant along the $a$ axis ($\varepsilon_a$) and the spontaneous ferroelectric polarization along the same direction ($P_a$) in fields of zero and 3 T applied along the $b$ axis.  Data in other field intensities were omitted for clarity.  We have confirmed the ferroelectric phase with transition temperatures of 5~K for cooling and 10~K for warming in a zero field.  
The lowest temperature we can reach is about 5~K, therefore, we can not directly detect $P_a$ for cooling.
However, the observed abrupt increase of $\varepsilon_a$ at 5~K for cooling strongly suggests the phase transition to the ferroelectric state.  Because we have observed the simultaneous change of $\varepsilon_a$ and $P_a$ as clearly shown in the warming scan. 
Reversal of ferroelectric polarization can also be observed by reversal of poling electric field.
Therefore, we have concluded that the ferroelectric phase along the $a$ axis exists below 5~K for cooling (10~K for warming) in a zero field~\cite{comments1}.

In a temperature range from 17~K to 24~K, one can immediately notice the broad peak of $P_a$ observed for only cooling, which is also an evidence of the ferroelectric phase along the $a$ axis.
For the cooling scan, a poling electric field of 500~kV/m was applied during the measurement to keep the sample in the single domain state of the ferroelectric phase~\cite{comments2}.
The origin of this upper ferroelectric phase is not clear at the present time.  We speculate that the upper ferroelectric phase is connected to the lower ferroelectric phase discussed above.  
The observed reentrant transition, i.e., ferroelectric/spiral-antiferromagnetic (FS) phase $ \rightarrow $ paraelectric/$A$-type antiferromagnetic (PA) one $ \rightarrow $ FS one, is quite intriguing, however, it is beyond the scope of this paper~\cite{comments3}. Hereafter, we restrict our consideration to the lower ferroelectric phase along the $a$ axis.

Let us move on to the magnetic field effect on the ferroelectricity.  As already reported in our previous reports~\cite{Noda1,Noda2,Noda3} and other groups'~\cite{Arima,Goto,Baier}, the magnetic field along the $b$ axis ($H_b$) stabilizes the ferroelectric phase. The observed ferroelectric transition temperatures are drastically increased by application of $H_b$, as shown in the Fig.~1 (See also Fig.~2~(a)).  On the other hand, magnetic fields along the $a$ and $c$ directions extremely suppress the ferroelectric phase. This is because these magnetic fields (especially $H_c$) stabilize the canted $A$-type antiferromagnetic and the resultant paraelectric (PA) phase, which competes and destabilizes the ferroelectric phase with non-collinear spiral spin structures (FS)~\cite{Arima,Goto,comments3}.  

\begin{figure}[h]

\includegraphics[width=160mm, keepaspectratio, clip]{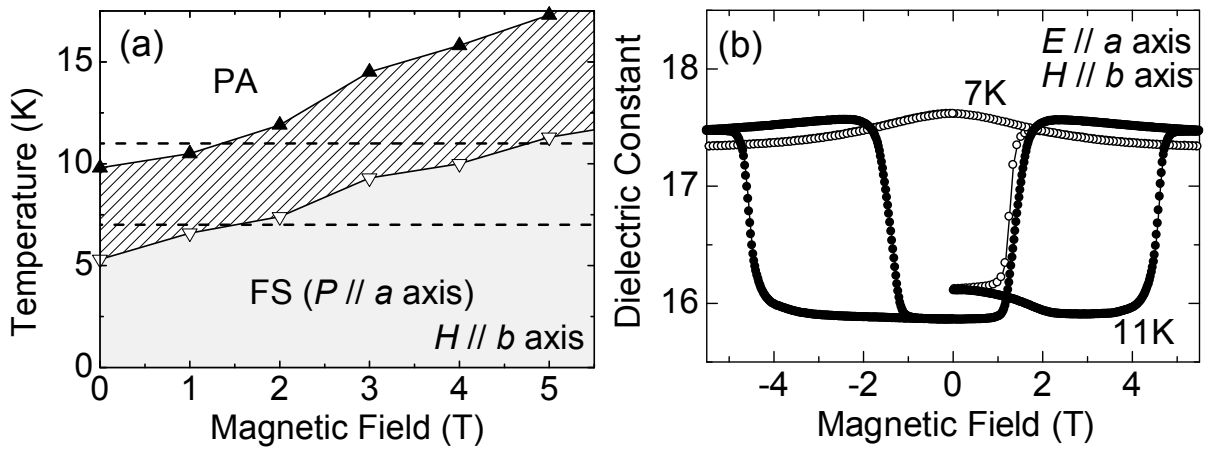}
\caption{\label{fig2} (a) Magnetoelectric phase diagram of GdMnO$_3$ in magnetic fields parallel to the $b$ axis.  
Hatched area denotes the hysteretic region between the \underline{p}araelectric/\underline{$A$}-type-antiferromagnetic (PA) and the \underline{f}erroelectric/\underline{s}piral-antiferromagnetic (FS) phases.\linebreak (b) Dielectric constant as a function of applied magnetic fields at fixed temperatures of 7~K  and 11~K\@.  Part of the trajectory of the measurements was shown by the dashed lines in (a).}

\end{figure}

We show in Fig.~2~(a) the magnetoelectric phase diagram of GdMnO$_3$ in the $H$-$T$ plane, which is determined by the measurements of $\varepsilon_a(T)$ in magnetic fields as shown in Fig.~1~(a). Open and solid triangles represent the ferroelectric transition temperatures for the cooling and warming runs, respectively. Hatched area indicates the hysteretic region reflecting the first order nature of the transition.  The large hysteretic behavior arises from the strong phase competition between the PA and FS states. This is in sharp contrast to the case of the second-order-like ferroelectric transition in TbMnO$_3$ and DyMnO$_3$, which reside far from the phase boundary between them and have a stable ferroelectric phase with spiral spin structure.

Using the large hysteretic region, we have successfully exemplified the phase switching from the PA phase to the FS one by application of magnetic fields along the $b$ axis.  Figure~2~(b) shows an example of such phase switching at two typical temperatures of 7~K and 11~K, which correspond to $within$ and $without$ the hysteretic region, respectively.  (See dashed lines in Fig.~2~(a).)  At a temperature of 11~K, the initial low $\varepsilon_a$ value at a zero field gradually decreases and suddenly jumps up to high $\varepsilon_a$ state at a field of 4.5~T for the field-increasing scan.  As easily noticed from Fig.~1~(a)~and~(b), low and high $\varepsilon_a$ sates correspond to the PA and FS ones, respectively.  The transition magnetic field of 4.5~T at 11~K is in good accordance with the lower phase boundary (open triangles) determined by the $\varepsilon_a(T)$ scans in $H_b$.
The high $\varepsilon_a$ value abruptly drops back to the low one again at 1.5~T for the field-decreasing scan, which indicates the transition from the FS phase back to the PA one again.
The transition field also agrees well with the upper phase boundary shown in Fig.~2~(a) (solid triangles). 
Furthermore, when magnetic fields were applied along the opposite direction ($H_b<0$), $\varepsilon_a(H_b)$ curves show the similar transition from the PA (FS) to the FS (PA) state at the same transition intensity of 4.5~T (1.5~T) for the field-increasing (-decreasing) scan, which is also accompanied by the magnetic-field hysteresis.
In this way, we can demonstrate the reversible switching between the PA and FS states. It should be noted that the ordered (ferroelectric) state is realized by application of magnetic fields. This is in contrast to the magnetic-field destruction of ordered state, e.g., the $P_a$ phase in GdMnO$_3$ is easily collapsed by the fields along the $a$ and $c$ directions. 

After initializing the sample heated well above the transition temperature, we have performed the same $\varepsilon_a(H_b)$ measurement at 7~K which is set $within$ the hysteretic region (the lower dashed line in Fig. 2 (a)). The initial low $\varepsilon_a$ value at a zero field sharply increases to the high one at a field of 1 T, which means the phase transition from the PA state to the FS one.  
The transition field was drastically reduced by lowering temperatures, as also seen in Fig.~2~(a). 
Once the sample has entered the FS state, the FS state subsists against the field-decreasing scan. 
Moreover, when magnetic fields were applied along the opposite direction like in the case of 11~K, the high $\varepsilon_a$ state is completely maintained throughout the field-increasing and -decreasing scans. This second case obviously demonstrates the persistent switching from the PA state to the FS one by application of $H_b$, which is in contrast to the case of reversible switching at 11~K\@.  To our knowledge, there has been no report on such persistent and reversible phase switching in other orthorhombic $R$MnO$_3$ systems.  This is partly because other $R$MnO$_3$ samples except for the present GdMnO$_3$ are far from the phase boundary between the PA and FS phases and are stabilized to the FS state $without$ the first order nature of the ferroelectric transition.

\section{Summary}

We have systematically investigated the dielectric properties as a function of temperature and magnetic fields for GdMnO$_3$, which is located in the vicinity of the phase boundary between the PA and FS phases.  We have deduced the magntoelectric phase diagram of GdMnO$_3$ in the $H$-$T$ plane, and have found the large hysteretic region due to the strong competition between the above two phases.  
Using the hysteretic region, we have successfully demonstrated the persistent and reversible phase switching between them by application of magnetic fields.  This is in sharp contrast to other $R$MnO$_3$ systems, which show only reversible phase switching.  We believe that these new findings provide the progress towards construction of a magnetic-recording erasable ferroelectric memory.

\ack

This work was partly supported by Grant-in-Aid for scientific research (C) from the Japan Society for Promotion of Science.


\begin{thebibliography}{9}
\bibitem{comments1} The ferroelectric ground state in a zero field is contradictory to other groups' reports, in which the ferroelectric phase can only exist in the fields along the $b$ axis.  This discrepancy arises from the slight difference of oxygen off-stoichomety.  We have checked the reproducibility of the ferroelectricity for several newly-grown samples, which is consistent with our previous works.
\bibitem{comments2} We have confirmed the ordinary pyroelectric current at an upper ferroelectric transition temperature of 24~K for warming without poling electric fields after the sample had been cooled to 20~K (not to the lowest temperature) with poling.
\bibitem{comments3} Detailed magnetic structure has not yet been determined because of the large absorption cross-section of Gd for neutrons.  
We assume that the ferroelectric phase has presumably the $ab$ spiral spin structures with the commensurate modulation wave vector. 
The magnetic modulation vector in the ferroelectric phase of GdMnO$_3$ is reported to be (0, 1/4, 1) (See ref.~1).
\end{thebibliography}
\end{document}